# Transporte Eletrônico no Semicondutor Carbeto de Silício na Fase 3C

**Amanda M. D. Corrêa, Clóves G. Rodrigues**

Escola Politécnica – Pontifícia Universidade Católica de Goiás (PUC Goiás)
Caixa Postal 86 – 74.605-010 – Goiânia – GO – Brazil

amandamdcorrea@gmail.com, cloves@pucgoias.edu.br

***Abstract.*** *In this work we study, theoretically, electronic mobility of the silicon carbide semiconductor in the 3C phase, named 3C-SiC. 3C-SiC has shown great potential for applications in extreme conditions. Thus, the study of the electronic mobility of this semiconductor is of great interest. In this work were theoretically deduced: drift velocity, displacement and mobility of the charge carriers in the n-type doped 3C-SiC semiconductor, subjected to a constant electric field. The dependence of these transport properties as a function of the intensity of the electric field and temperature was analyzed. For this, a differential equation of motion was used with a source term (low intensity electric fields) and a term of resistance to movement (electrical resistance).*

***Resumo.*** *Neste artigo foi determinada teoricamente a mobilidade eletrônica do semicondutor carbeto de silício na fase 3C, chamado de 3C-SiC. O 3C-SiC tem mostrado uma grande potencialidade para aplicações em condições extremas. Assim o estudo da mobilidade eletrônica deste semicondutor é de grande interesse. Neste trabalho foi deduzido teoricamente a velocidade de deriva, o deslocamento e a mobilidade dos portadores de carga no semicondutor 3C-SiC dopado tipo n e submetido a um campo elétrico constante sendo analisada a dependência destas propriedades com a intensidade do campo elétrico e temperatura. Para tanto, foi utilizada uma equação diferencial de movimento com um termo de fonte (devido a campos elétricos) e um termo de resistência ao movimento (resistência elétrica).*

## 1. Introdução

O semicondutor é um material com propriedades elétricas entre as dos condutores e as dos isolantes. Entre os vários materiais que existem, a propriedade física que apresenta a maior variação é justamente a condutividade elétrica dos materiais. Esta pode variar de $10^{-18} \Omega^{-1} m^{-1}$ (quartzo) a $10^{8} \Omega^{-1} m^{-1}$ (prata, cobre), ou várias ordens de grandeza maiores que isso, no caso de supercondutores. Semicondutores são materiais para os quais a uma temperatura de zero Kelvin, a banda de valência está totalmente preenchida e a banda de condução totalmente vazia, funcionando nessa condição como isolantes. Uma maneira de se produzir elétrons livres para condução em um semicondutor é através da adição de impurezas, processo que é chamado de *dopagem* [Ashcroft, 1976].

Um semicondutor de interesse atual é o carbeto de silício (SiC), o qual possui várias utilidades comerciais. Inúmeras aplicações eletrônicas e optoeletrônicas tem sido propostas com base nas propriedades eletrônicas e ópticas do SiC, como: microestruturas, dispositivos opto-eletrônicos, eletrônicos de alta temperatura, eletrônicos rígidos de radiação, e dispositivos de alta potência e alta frequência [Rodrigues, 2010]. O SiC possui vários politipos sendo os principais o 3C-SiC, o 4H-SiC e o 6H-SiC.

Este trabalho é dirigido ao semicondutor Carbeto de Silício (SiC) na fase 3C (veja Fig. 1), o qual é também chamado de 3C-SiC, onde o C- indica que sua simetria é cúbica e o 3 refere-se à forma de empilhamento atômico. Neste artigo foi deduzido teoricamente a velocidade de deriva, o deslocamento e a mobilidade dos portadores de carga (elétrons de condução) no semicondutor 3C-SiC dopado tipo *n*, submetido a um campo elétrico constante. As propriedades de transporte das fases 4H-SiC e 6H-SiC já foram estudadas e os resultados publicados [Vasconcelos, 2019a; Vasconcelos, 2019b; Ferracioli, 2020].

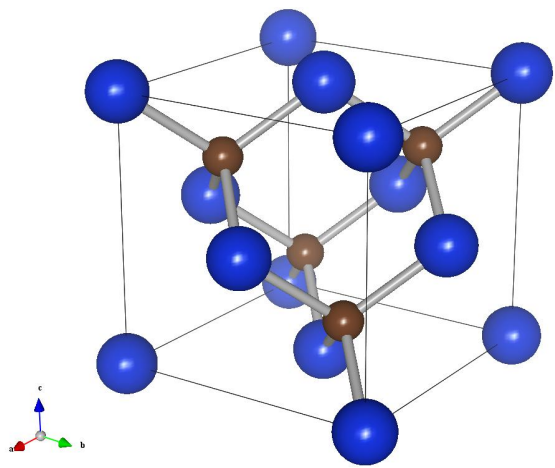

**Figura 1. Estrutura cristalina do 3C-SiC.**

## 2. Descrição do Movimento do Elétron de Condução no Semicondutor Submetido a Campo Elétrico

Neste artigo considera-se um semicondutor dopado tipo *n* com uma concentração de impurezas menor que $10^{18}$ cm$^{-3}$. Nesta condição não é necessário levar em consideração o espalhamento dos portadores de carga (elétrons) pelas impurezas [Rodrigues, 2009]. O movimento dos portadores será governado por uma força externa $\vec{F}_{ext}$ e por uma força de resistência ao movimento $\vec{f}$. Para descrever tal situação foi utilizada uma equação semi-clássica baseada na segunda lei de Newton [Kittel, 2006], ou seja:

$$\frac{d\vec{P}}{dt} = \sum \vec{F} = \vec{F}_{ext} + \vec{f}. \quad (1)$$

Na Equação (1) será aplicada a forma quântica do momento total $\vec{P}$ e a expressão quântica para $\vec{f}$ [Rodrigues, 2020]. Para utilizar a forma semi-clássica da Segunda Lei de Newton, é necessário quantizar o momento $\vec{P}$ do elétron, que neste caso resulta em [Rodrigues, 2000]:

$$P(t) = m_e^* v(t), \quad (2)$$

onde $m_e^*$ é a massa efetiva dos portadores. A massa efetiva dos portadores é uma massa que permite usar a segunda lei de Newton da Física Clássica. Isto porque a massa efetiva engloba os efeitos quânticos do potencial interno da rede cristalina sobre os portadores de carga. Substituindo a Eq. (2) na Eq. (1) tem-se:

$$m_e^* \frac{dv(t)}{dt} = \vec{F}_{ext} + \vec{f}. \tag{3}$$

A força $\vec{F}_{ext}$ surge pela aplicação de um campo elétrico ao material semicondutor, portanto esta será uma força elétrica $\vec{F}_{el}$ [Reitz, 1982]. Por ser o campo elétrico um campo vetorial, temos associado a cada ponto do espaço um vetor campo elétrico. Assim a força elétrica $\vec{F}_{el}$ que atua sobre um elétron tem a mesma direção e sentido oposto à orientação do vetor campo elétrico que atua sobre o elétron, conforme ilustrado na Fig. 2 (os vetores não estão em escala).

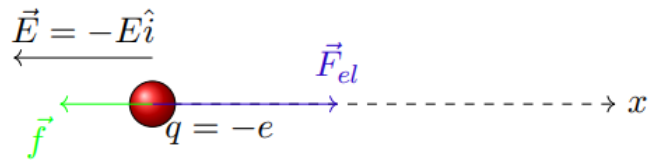

**Figura 2. Esquema das forças que atuam sobre o elétron quando um campo elétrico, aplicado na direção do eixo X, passa a atuar sobre o elétron.**

A força elétrica atuando num elétron de condução com carga $q = -e$ é por definição [Reitz, 1982]:

$$\vec{F}_{el} = q\,\vec{E} = (-e)(-E\hat{\imath}) = eE\hat{\imath}, \tag{4}$$

onde $e$ é a carga elementar do elétron e $E$ é o módulo do campo elétrico aplicado.

Como o elétron está se movendo dentro de um meio material, há uma força de resistência $\vec{f}$ ao movimento do elétron que depende das características do meio material. Considera-se que essa força é proporcional à velocidade $\vec{v}$ do elétron de condução [Rodrigues, 2000], ou seja:

$$\vec{f} = -\alpha\,\vec{v}, \tag{5}$$

onde $\alpha$ é um parâmetro que está associado à resistividade elétrica do semicondutor [Rodrigues, 2000]. Substituindo as Eqs. (4) e (5) na Eq. (3), tem-se:

$$m_e^* \frac{d\vec{v}}{dt} = e\,E\hat{\imath} - \alpha\vec{v}. \tag{6}$$

Pelo fato do movimento ocorrer somente ao longo do eixo $x$ pode-se escrever a equação anterior sem a simbologia vetorial da seguinte maneira:

$$m_e^* \frac{dv}{dt} = eE - \alpha v, \tag{7}$$

sendo $m_e^*$, $e$, $E$ e $\alpha$ quantidades positivas.

A Equação (7) pode ser solucionada de forma exata por meio de técnicas de integração [Rodrigues, 2017], obtendo-se:

$$v(t) = \frac{eE}{\alpha}\left[1 - e^{-\frac{\alpha\,t}{m_e^*}}\right]. \tag{8}$$

A partir da Eq. (8) determina-se por integração a equação da posição do elétron, a saber:

$$x(t) = \frac{eE}{\alpha}\left[t + \frac{m_e^*}{\alpha}(1 - e^{\frac{\alpha t}{m_e^*}})\right]. \qquad (9)$$

A expressão do parâmetro $\alpha$ que aparece nas Eqs. (8) e (9) é dada no Apêndice A.

## 3. Resultados para o 3C-SiC

Serão utilizadas as Eqs. (8) e (9), da seção anterior, para plotar a velocidade e a posição do elétron de condução em função do tempo no semicondutor 3C-SiC. Os parâmetros do semicondutor 3C-SiC utilizados nos cálculos numéricos estão disponíveis na Tabela A.1 do Apêndice A. Para obter numericamente a velocidade e a posição dos elétrons de condução em função do tempo foi utilizado o programa computacional Mathematica versão 11.0 [Wolfram, 2022].

A Fig. 3 apresenta a evolução temporal da velocidade dos elétrons de condução para três valores de campos elétricos aplicados: 1 kV/cm, 2 kV/cm e 3 kV/cm. A temperatura ambiente considerada foi de 300 K. Observando a Fig. 3 nota-se que a velocidade do elétron aumenta com o valor do campo elétrico aplicado, e que após um intervalo de tempo de aproximadamente 1 ps, tende a se tornar constante (estado estacionário).

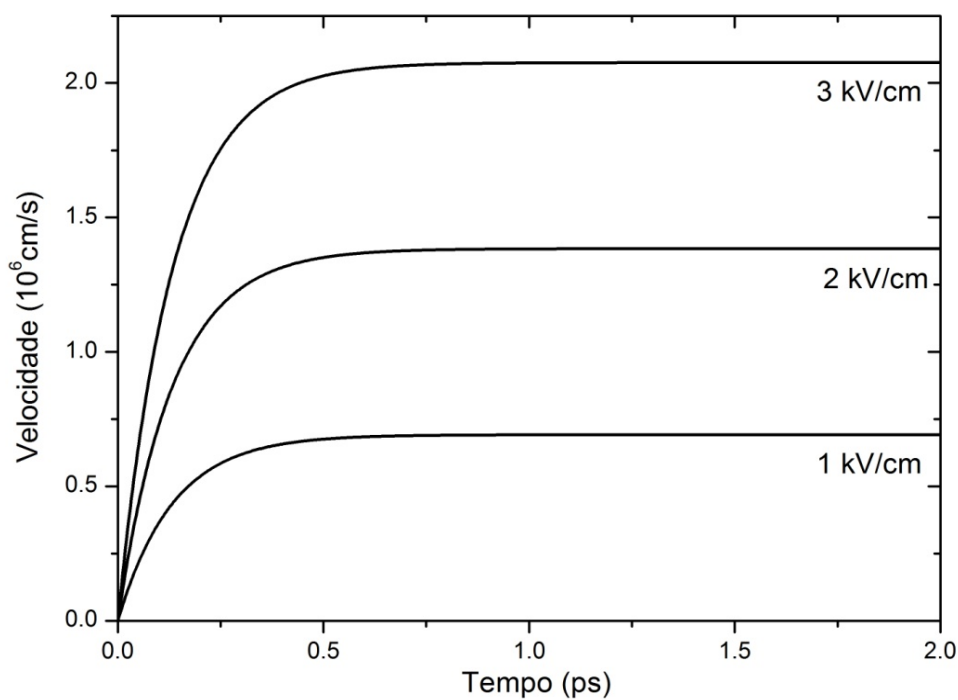

**Figura 3. Velocidade do elétron em função do tempo no semicondutor 3C-SiC.**

Utilizando a Eq. (9) da seção anterior, determina-se a evolução temporal da posição do elétron de condução no semicondutor 3C-SiC para três valores de campos elétricos aplicados (1, 2 e 3 kV/cm) e para uma temperatura ambiente de 300 K, conforme ilustrado na Fig. 4. Em um intervalo de tempo de 2 ps os deslocamentos foram de 0,013 μm, 0,026 μm, e 0,039 μm para os campos elétricos de 1, 2 e 3 kV/cm, respectivamente.

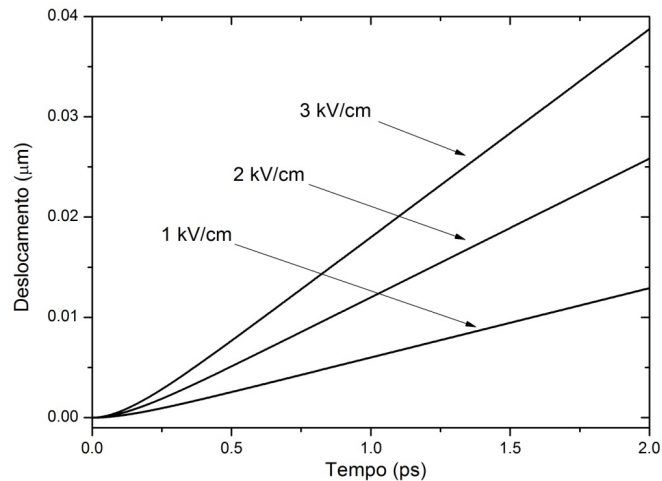

**Figura 4. Posição do elétron em função do tempo no semicondutor 3C-SiC.**

No estado estacionário a velocidade de deslocamento do elétron torna-se constante, sendo chamada de velocidade estacionária ($v_{est}$). Isto acontece quando a força de resistência ao movimento $f$ torna-se igual em módulo à força externa aplicada, e conforme a Eq. (7):

$$0 = eE - \alpha v_{est} \Rightarrow v_{est} = \frac{eE}{\alpha}, \qquad (10)$$

lembrando que o parâmetro $\alpha$ é dado no Apêndice A. Utilizando os dados do semicondutor 3C-SiC, apresenta-se na Fig. 5 o comportamento da velocidade de deriva do elétron no estado estacionário em função do campo elétrico, fornecido pela Eq. (10), utilizando uma temperatura ambiente de 300 K. Verifica-se que no intervalo de 0 a 3 kV/cm a velocidade estacionária dos elétrons de condução no semicondutor 3C-SiC aumenta de forma linear com o aumento da intensidade do campo elétrico aplicado, exibindo um comportamento ôhmico nesta faixa de campo elétrico aplicado.

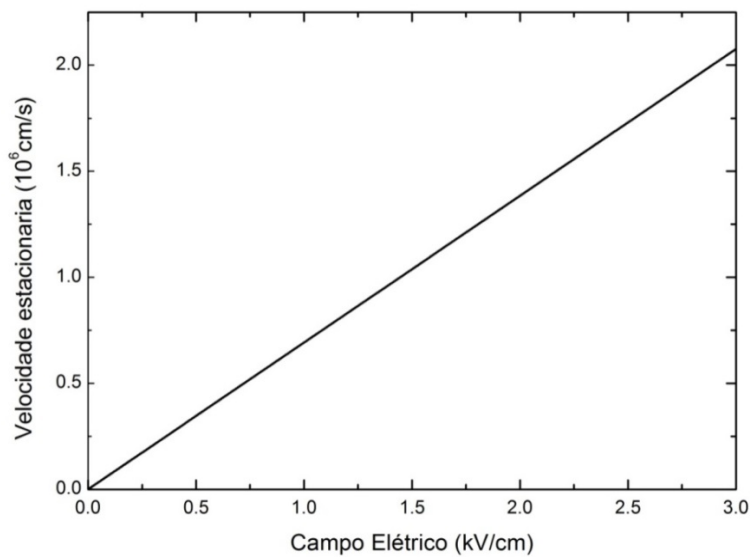

**Figura 5. Velocidade estacionária dos elétrons de condução em função da intensidade do campo elétrico.**

As Figuras 3, 4 e 5 foram produzidas utilizando-se uma temperatura ambiente fixa, com valor de 300 K. No entanto, o parâmetro $\alpha$ possui uma dependência com a temperatura da rede cristalina (veja Apêndice A). A Fig. 6 ilustra o comportamento da velocidade estacionária dos portadores de carga em função da temperatura da rede cristalina para o carbeto de silício na fase 3C. Com o aumento da temperatura, aumenta-se a vibração da rede cristalina, aumentando a possibilidade de colisão dos portadores de carga com os átomos da rede. Dessa forma, os portadores de carga perdem parte de sua energia de movimento, fazendo com que a intensidade da velocidade dos elétrons de condução diminua com o aumento da temperatura.

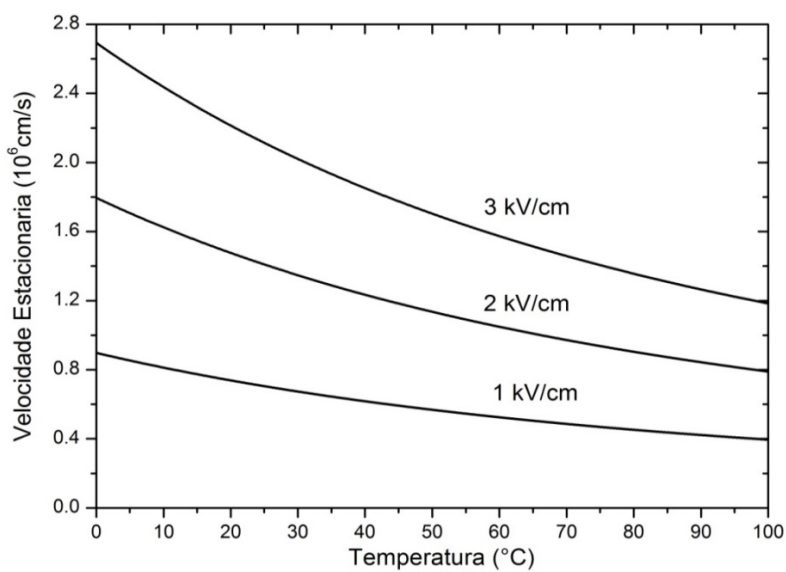

**Figura 6. Velocidade estacionária dos portadores de carga elétrica em função da temperatura da rede cristalina.**

Apesar da mobilidade eletrônica não depender da intensidade do campo elétrico nas condições aqui estudadas (baixos valores de campo elétrico – regime ôhmico), ela é influenciada pela temperatura. De forma análoga à velocidade, a mobilidade eletrônica também diminui com o aumento da temperatura, como mostra a Fig. 7. Nota-se uma redução de aproximadamente 56% na mobilidade de 0 a 100ºC.

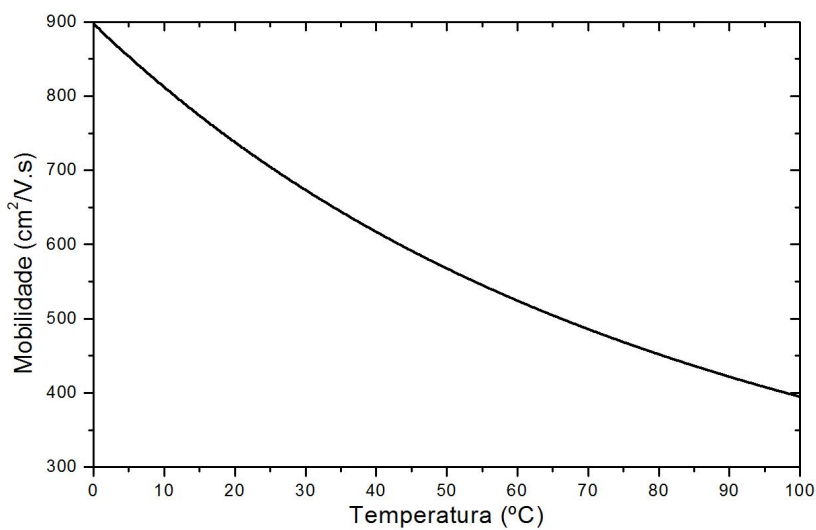

**Figura 7. Mobilidade eletrônica em função da temperatura.**

A energia cinética por elétron, $K_{cin}$, no estado estacionário é definida por: $K_{cin} = m_e^* v_{est}^2/2$. Lembrando que a velocidade do elétron no estado estacionário é dada por $v_{est} = eE/\alpha$ (vide Eq. 10), a energia cinética por elétron assume a forma: $K_{cin} = (m_e^* e^2/2\alpha^2)E^2$, ou seja, a energia cinética possui uma dependência quadrática com o módulo do campo elétrico $E$. A Fig. 8 mostra este comportamento, onde é plotada a energia cinética por elétron em função da intensidade do campo elétrico aplicado, variando-se a intensidade do campo elétrico de 0 até 3 kV/cm. A unidade utilizada para a energia cinética está em "mili elétron-Volts", a qual é uma unidade de energia mais adequada para o estudo do transporte eletrônico em semicondutores. Nota-se pela Fig. 8 que esta energia é da ordem de sub meV.

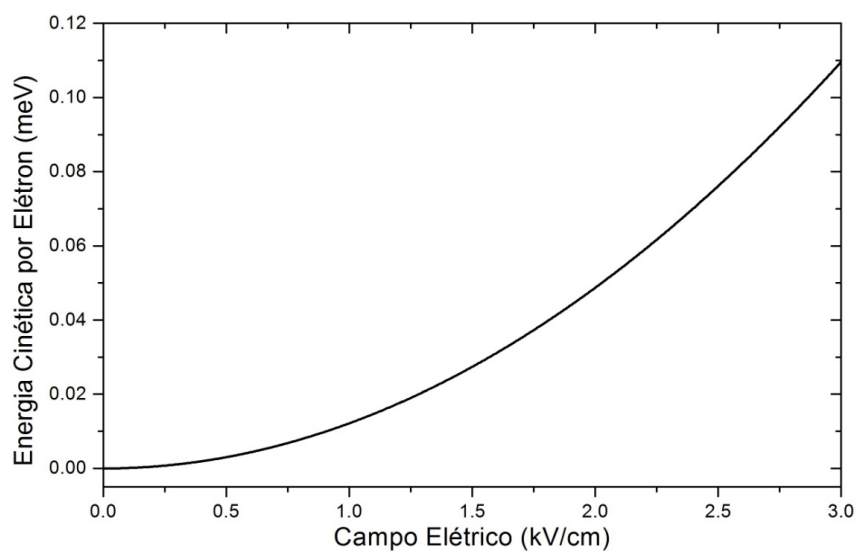

**Figura 8. Energia cinética por elétron em função do campo elétrico aplicado.**

Uma importante propriedade dos materiais semicondutores é a mobilidade eletrônica $\mu$, definida como [Madelung, 1996]:

$$\mu = \frac{v_{est}}{E}, \qquad (11)$$

ou seja, a razão entre a velocidade dos elétrons de condução no estado estacionário e o campo elétrico aplicado. A mobilidade $\mu$ pode, então, ser obtida pela Fig. 5 calculando-se a declividade da reta. Assim obtemos para o 3C-SiC uma mobilidade $\mu = 693$ cm$^2$/V.s.

A Tabela 1 mostra uma comparação da mobilidade eletrônica do 3C-SiC em relação aos dois semicondutores mais utilizados comercialmente na atualidade. Nota-se que a mobilidade do Ge e do Si são bem superiores à do 3C-SiC, no entanto, estes dois semicondutores tem suas aplicações limitadas em condições extremas (altas temperaturas, altas tensões, altas pressões, altos níveis de radiação eletromagnética) o que não acontece com o 3C-SiC devido à sua alta rigidez mecânica e ao seu gap largo. Quanto à comparação com os resultados experimentais, existe uma concordância relativamente boa do resultado aqui obtido com os valores experimentais. Nos trabalhos experimentais da Tabela 1, Piluso et al. (2011) estudam a mobilidade elétrica, utilizando espectroscopia micro-Raman, em filmes finos de 3C-SiC, dopados tipo $n$ e cultivados em substratos orientados de silício. Nishino et al. (1987) estudam a mobilidade Hall de monocristais de SiC cúbico cultivados heteroepitaxialmente por deposição química de vapor (CVD) usando um sistema $SiH_4$-$C_3H_8$-$H_2$ em substratos de silício, enquanto Roschke (2001) apresenta uma compilação de dados experimentais utilizados em seu artigo para um modelo de aproximação de difusão de deriva baseado na mobilidade eletrônica experimental.

**Tabela 1. Mobilidade Eletrônica, à temperatura ambiente, em cm²/V.s**

| Semicondutor | Mobilidade | Método | Referência | Diferença |
|---|---|---|---|---|
| Si | 1350 | experimental | [Kitell, 2006] | - |
| Ge | 3600 | experimental | [Kitell, 2006] | - |
| 3C-SiC | 750 | experimental | [Nishino, 1987] | 7,6% |
| 3C-SiC | 693 | teórico | neste trabalho | - |
| 3C-SiC | 650 | experimental | [Roschke, 2001] | 6,2% |
| 3C-SiC | 570 | experimental | [Piluso, 2011] | 17,7% |

Uma boa concordância entre experimento e teoria é fundamental para a validação do método teórico empregado. As variações nos valores experimentais para a mobilidade acontecem devido a uma não padronização das amostras utilizadas e também com imperfeições que ocorrem no crescimento do cristal como: defeitos na rede cristalina, deslocamentos, vacâncias, impurezas indesejadas, etc. Outro ponto a ser reverenciado é a dificuldade e imprecisão das constantes características do semicondutor 3C-SiC utilizadas nos cálculos numéricos (veja Tabela A.1 no Anexo A), as quais influenciam diretamente nos resultados obtidos.

## 4. Comentários Finais

O semicondutor 3C-SiC possui uma grande potencialidade para aplicações em condições extremas de operação, tais como: altas temperaturas, altas tensões mecânicas, elevados campos elétricos, altas tensões elétricas, altos níveis de radiação. Isto se deve em grande parte ao seu grande gap de energia e alta rigidez mecânica. Assim o estudo da mobilidade eletrônica deste semicondutor é de grande interesse para a indústria de semicondutores.

Neste trabalho foi deduzido teoricamente a velocidade de deriva, o deslocamento e a mobilidade dos portadores de carga no semicondutor 3C-SiC dopado tipo *n* e submetido a um campo elétrico constante. O método utilizado foi a montagem de uma equação diferencial de movimento com um termo de fonte, o qual representa o campo elétrico externo aplicado, e um termo de resistência ao movimento, o qual representa matematicamente as colisões que os portadores de carga sofrem ao se deslocarem pela rede cristalina do semicondutor na direção do campo elétrico aplicado. Estas quantidades (velocidade, deslocamento e mobilidade) são comumente chamadas de "propriedades de transporte" e são de extrema importância do ponto de vista de aplicações em dispositivos eletrônicos. Essas informações das propriedades de transporte em semicondutores possibilitam construir dispositivos com tamanhos específicos que podem levar ao dimensionamento mais eficiente dos custos de produção destes dispositivos. A dependência destas propriedades de transporte em função da intensidade do campo elétrico e da temperatura para o 3C-SiC foi analisada e verificou-se uma grande diminuição da mobilidade com o aumento da temperatura.

Quanto à comparação com resultados experimentais, existe uma concordância relativamente boa do resultado aqui obtido com os valores experimentais encontrados na literatura.

## Apêndice. O Parâmetro $\alpha$

Nas Equações (8) e (9) o parâmetro $\alpha$ é dado por [Rodrigues, 2000]:

$$\alpha = \frac{e^2 \omega^2 \gamma}{3} \sqrt{\frac{2(m_e^*)^2}{\pi (k_B T)^3}} \frac{2 e^z K_1(z)}{e^{2z} - 1} + \frac{(2\pi)^4 E_{1e}^2 \sqrt{(m_e^*)^5 (2 k_B T / \pi)^3}}{3 h^4 \rho (v_s)^2}, \quad (A.1)$$

onde

$$\Gamma = \frac{1}{\varepsilon_\infty} - \frac{1}{\varepsilon_0}, \quad (A.2)$$

$$z = \frac{h \omega}{4 \pi k_B T}, \quad (A.3)$$

sendo $e$ a carga elementar do elétron, $\omega$ a frequência dos fônons ópticos, $m_e^*$ a massa efetiva do elétron de condução, $\varepsilon_0$ a constante eletrostática estática, $\varepsilon_\infty$ a constante eletrostática de alta frequência, $T$ a temperatura da rede cristalina, $K_1(z)$ a função modificada de Bessel de segunda espécie com argumento $z$, $k_B$ a constante de Boltzmann, $h$ a constante de Planck, $\rho$ a densidade do material, $v_s$ a velocidade do som no material e $E_{1e}$ o potencial de deformação acústico. Os valores característicos destes parâmetros para o semicondutor 3C-SiC e os valores das constantes físicas utilizadas nos cálculos numéricos estão dispostos nas Tabelas A.1 e A.2, respectivamente.

**Tabela A.1. Parâmetros Característicos do 3C-SiC.**

| Parâmetro | Símbolo | Valor |
|---|---|---|
| Frequência dos fônons ópticos | $\omega$ | 1,82x10$^{14}$s$^{-1}$ [Bellotti, 1999] |
| Massa efetiva do elétron | $m_e^*$ | 0.346 $m_0$ [Weng, 1997] |
| Constante eletrostática estática | $\varepsilon_0$ | 9,72 [Patrick, 1970] |
| Constante eletrostática de alta frequência | $\varepsilon_\infty$ | 6,52 [Patrick, 1970] |
| Densidade do material | $\rho$ | 3,21 g/cm$^3$ [Harris, 1995] |
| Velocidade do som no material | $v_s$ | 8,33x10$^5$ cm/s [Bellotti, 1999] |
| Potencial de deformação acústico | $E_{1e}$ | 22 eV [Yamanaka, 1987] |

**Tabela A.2. Constantes Físicas Utilizadas [Ashcroft, 1976].**

| Constante | Símbolo | Valor |
|---|---|---|
| Carga elementar do elétron | $e$ | $1{,}602 \times 10^{-12}$ C |
| Constante de Boltzmann | $k_B$ | $1{,}381 \times 10^{-23}$ J/K |
| Constante de Planck | $h$ | $6{,}626 \times 10^{-34}$ J.s |
| Massa de repouso do elétron | $m_0$ | $9{,}109 \times 10^{-34}$ g |